\documentclass[a4paper]{article}
\usepackage[a4paper, total={6in, 8.5in}]{geometry}
\usepackage[english]{babel}
\usepackage[utf8]{inputenc}
\usepackage{amsmath}
\usepackage{amssymb}
\usepackage{graphicx}
\usepackage{xurl}
\usepackage[hidelinks]{hyperref}

\usepackage[colorinlistoftodos]{todonotes}

\usepackage[
    backend=biber,
    style=numeric-comp,
    sorting=none,
    doi=true,
    url=false,
    eprint=false
]{biblatex}
\usepackage{authblk}  
\newif\ifmain
\newif\ifsupp

\maintrue
\supptrue

\usepackage[utf8]{inputenc}
\usepackage{newunicodechar}
\usepackage{lineno}
\newunicodechar{ }{\,}   
\newunicodechar{–}{--}   
\newunicodechar{—}{---}  
\newunicodechar{’}{'}    
\newunicodechar{‘}{`}    
\newunicodechar{“}{``}   
\newunicodechar{”}{''}
\newunicodechar{γ}{$\gamma$}

\title{A Dynamical Blueprint for Brain State Organization}

\author[1,2]{Kateryna Nechyporenko}
\author[2]{Peter Ashwin}
\author[1,2]{Krasimira Tsaneva-Atanasova}

\affil[1]{Living Systems Institute, University of Exeter, Stocker Road, Exeter EX4 4PY, UK}
\affil[2]{Department of Mathematics and Statistics, University of Exeter, Harrison Building, Exeter EX4 4QF, UK}

\date{}

\usepackage{float}

\begin{document}

\ifmain


\maketitle
\noindent Kateryna Nechyporenko is the first author. Correspondence should be addressed to Kateryna Nechyporenko (k.nechyporenko@exeter.ac.uk).

\section*{Abstract}
The brain is not static: neuronal networks shift between contrasting modes of activity, alternating between active and quiescent regimes known as up and down states. Together with rhythmic oscillations, such modes of activity are fundamental to perception, memory, and information processing. However, the dynamical principles underlying the diverse repertoire of activity patterns and their transitions remain poorly understood. Here, we identify a geometric structure that governs dynamic states emergence and organizes neuronal networks transitions. We derive the conditions for its existence and demonstrate that it emerges robustly across canonical models of neuronal population dynamics. Near this organizing center, switches between oscillations, bistability and up and down states are orchestrated by the excitation-inhibition balance in the neuronal network. Thus, we show that excitation and inhibition do not simply modulate network activity but define the dynamical landscape from which distinct brain states emerge. We also consider neuron-astrocyte interactions and reveal how astrocytes can tune excitatory-inhibitory balance, therefore modulating the transitions between neuronal activity regimes. Overall, our results identify a general dynamical blueprint underlying the emergence, organization, and control of brain states. 

\noindent\textbf{Teaser:} Excitation-inhibition balance organizes brain states through a shared dynamical blueprint.

\section*{Introduction}
Neuronal network activity often exhibits transitions between distinct, experimentally observable regimes, such as up and down states, corresponding to periods of sustained firing and relative quiescence \cite{tukker_up_2020, brinkman_metastable_2022, jercog_up-down_2017,levenstein_nrem_2019}. Understanding how the transitions between these states emerge is central to explaining how neural circuits dynamically regulate information processing. Dynamical systems theory provides a framework for studying the mechanisms underlying the state transitions. However, despite decades of work, there is still no unifying principle explaining how such transitions are organized across neuronal systems, limiting our ability to determine whether similar activity patterns, that are observed across neuronal systems, are driven by common mechanisms.

Such state transitions are often associated with metastable dynamics, in which neural activity transiently occupies semi-stable dynamical regimes before switching to others. In non-chaotic systems, metastable dynamics can arise through mechanisms such as multistability, noise- or perturbation-driven transitions between coexisting attractors, and sequential trajectories shaped by invariant structures, including slow manifolds and trajectories such as heteroclinic connections \cite{rossi_dynamical_2025}. While multistability provides robustness through sustained activity states \cite{holcman_emergence_2006}, heteroclinic dynamics supports order and timing through sequential transitions \cite{laurent_odor_2001,afraimovich_origin_2004}. In addition, experimental observations of natural non-rapid eye movement (NREM) sleep are explained by excitable dynamics, in which a stable state is intermittently interrupted by fluctuation-induced transient events \cite{levenstein_nrem_2019}. Hence, rather than being constrained to a single mode of operation, neural systems may exploit the interplay between complementary dynamical mechanisms to balance robustness, flexibility and temporal organization in cognitive processes. Deco and Rolls proposed a model in which sequential memory emerges from transitions between attractor states driven by short-term synaptic depression, which destabilizes persistent activity and induces ordered switching between neural representations \cite{deco_sequential_2005}. Related theoretical frameworks suggest that hippocampal-prefrontal circuits may combine persistent and sequential dynamics to support temporal organization of cognitive processes \cite{friston_functional_2016}. Winnerless competition further illustrates this interplay, where non-symmetric inhibition destabilizes winner-take-all equilibria, giving rise to sequential switching organized by heteroclinic dynamics \cite{afraimovich_heteroclinic_2004, rabinovich_dynamical_2006}. Despite these advances, there is currently no general dynamical systems theory that unifies bistable and sequential regimes, leaving the relationship between these mechanisms an open question.

Phenomenological mean-field models provide a reduced description of the collective dynamics of neuronal populations and are widely used to study the emergence of distinct neuronal activity regimes. In doing so, they offer a framework for relating observed brain activity to the underlying physiological mechanisms \cite{dunstan_neural_2025,sanchez-rodriguez_personalized_2024,van_nifterick_multiscale_2022, ginsberg_mechanisms_2025}. Canonical examples of phenomenological mean-field models include the Wilson-Cowan framework, which describes interactions between excitatory and inhibitory populations through firing rate dynamics \cite{wilson_excitatory_1972, wilson_mathematical_1973}. In contrast, models based on the Tsodyks-Markram framework shift the focus to the synaptic level, modeling short-term plasticity through neurotransmitter depletion (synaptic depression) and recovery (synaptic facilitation) \cite{tsodyks_neural_1997,tsodyks_neural_1998,mongillo_synaptic_2008,markram_differential_1998,loebel_multiquantal_2009}. Another widely used class of phenomenological mean-field models is based on the Jansen-Rit framework, designed to reproduce macroscopic signals including electroencephalogram and local field potentials \cite{jansen_electroencephalogram_1995,robinson_propagation_1997,wendling_relevance_2000,liley_bifurcations_2010,liley_spatially_2002,freeman_nonlinear_1979}. The Jansen-Rit model is structured as a network of pyramidal neurons receiving input from both inhibitory and excitatory interneuron populations, with model equations representing a realization of the Wilson-Cowan framework in higher dimensions \cite{coombes_neurodynamics_2023}. Although these frameworks differ in their physiological interpretation and level of description, they share a common mathematical foundation as coupled nonlinear dynamical systems capable of generating a rich repertoire of activity regimes. While the dynamical mechanisms underlying many of the individual regimes are well understood, considerably less is known about the organizing principles that link them and govern transitions between them. Identifying such a blueprint is essential for understanding how physiological changes at the cellular, synaptic, or network level can reshape large-scale neuronal dynamics.

Motivated by this gap, we aim to identify a unifying dynamical structure that governs transitions between distinct activity regimes across neuronal systems. To this end, we introduce a double saddle-node bifurcation on invariant circle (SNIC) organizing center, SNIC$^2$, a low-dimensional dynamical blueprint that unifies bistable, excitable, and sequential regimes while organizing transitions between them. We derive the necessary conditions for its emergence and demonstrate that this phenomenon arises across canonical mean-field models, such as Wilson–Cowan, Tsodyks–Markram, and Jansen–Rit. Furthermore, by extending a Wilson–Cowan framework to include astrocyte-mediated interactions, we show how astrocytes can modulate transitions between network activity states. Together, our results identify a shared dynamical architecture through which excitation–inhibition balance organizes state transitions across neuronal systems. More broadly, SNIC$^2$ provides a model-based dynamical framework for relating changes in effective excitatory and inhibitory drive to qualitative reorganizations of network activity.

\section*{Results}
\subsection*{Conserved phase-plane organization in mean-field neuronal network models}\label{part1}
Phenomenological mean-field models of brain dynamics can exhibit a range of activity regimes, including high-activity states (up states), quiescence (down states), oscillations (limit cycle solutions), and bistability (coexistence of up and down states). We identify a codimension-two dynamical structure, termed SNIC$^2$, that arises at the intersection of two SNIC curves, corresponding to the simultaneous occurrence of two saddle-node bifurcations on invariant circles (\autoref{fig:1}\textbf{A}). At this point, the system exhibits a heteroclinic loop connecting two non-hyperbolic saddle equilibria (saddle-nodes). The proposed name, SNIC$^2$, reflects the fact that an organizing center emerges from the intersection of two distinct saddle-node on invariant circle (SNIC) curves. Homburg and Sandstede \cite{homburg_chapter_2010} noted that a heteroclinic cycle involving two saddle-nodes becomes a codimension-two object when it forms a hyperbolic invariant circle, corresponding to the SNIC$^2$ configuration identified here (see \autoref{fig:1}\textbf{A}, pink region).

\begin{figure*}[!tbhp]
\centering
\includegraphics[width=.8\linewidth]{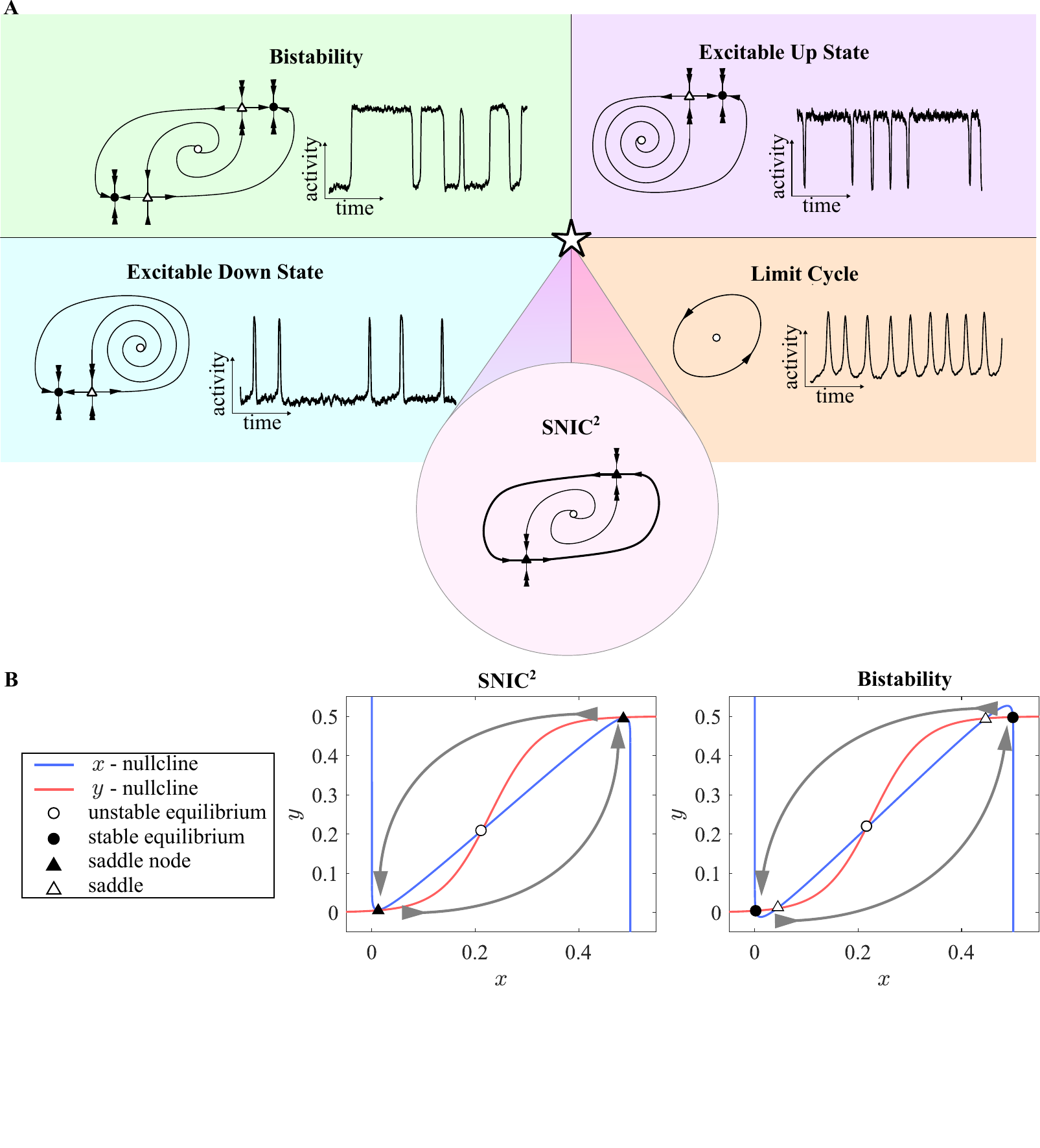}
\caption{\textbf{A} Sketch of the transition between different dynamic states via SNIC$^2$ organzing center and the associated with them time series. The dynamic transitions from an excitable down state (blue) to a bistable regime (green) to an excitable up state (purple) to a limit cycle (orange). The organizing center of all of the transitions is SNIC$^2$ - heteroclinic cycle between two saddle nodes (pink). Double arrow signifies the direction of the strong stable manifold, while white triangle, black triangle, white circle and black circle describe hyperbolic saddle, saddle node, unstable equilibrium and a stable equilibrium, respectively. \textbf{B} An example of the nullcline configuration that gives rise to the SNIC$^2$ organizing center (left), along with the configuration observed when the system transitions from SNIC$^2$ to bistability (right) via a simultaneous splitting of both saddle nodes into a saddle and a node.}
\label{fig:1} 
\end{figure*}

To characterize the geometry of the SNIC$^2$ phenomenon and establish the conditions under which it can arise, we derive the necessary requirements for its existence in a low dimensional setting. Here, these conditions can be formulated as geometric constraints on the relative arrangement of the nullclines in the phase plane. Consider a two-dimensional autonomous dynamical system defined on an open domain $\Omega \subseteq \mathbb{R}^2$:
\begin{align}
    \dot x &= f(x,y), \label{gen:eq1} \\
    \dot y &= g(x,y), \label{gen:eq2}
\end{align}
where \( f, g : \Omega \to \mathbb{R} \) are nonlinear smooth functions of class \( C^{k} \), for some \( k \geq 2 \).
We define the nullclines of the system as the zero-level sets:
\begin{align}
    \mathcal{N}_f = \{ (x, y) \in \Omega : f(x, y) = 0 \}, \nonumber \\
    \mathcal{N}_g = \{ (x, y) \in \Omega : g(x, y) = 0 \}. \nonumber
\end{align}

\paragraph{Necessary conditions for SNIC$^2$ in a two-dimensional system.} We now state the necessary conditions (\textbf{NC}s) for the existence of two non-hyperbolic saddle points that can form a heteroclinic cycle. 
\begin{itemize}
    \item \textbf{(NC1)} \textit{Critical points of a nullcline function}:
    There exists a  $C^{k}$ (for some \( k \geq 2 \)) function $f(x,y)$ such that for some $y = y_{\min}$ and $y = y_{\max}$, the function $f(x) := f(x, y)$ satisfies:
    $\exists\; x_{\min}, x_{\max} \in \mathbb{R}$ with $x_{\min} < x_{\max}$, such that:
        $$
        f_x(x_{\min}, y_{\min}) = 0, \quad f_{xx}(x_{\min}, y_{\min}) > 0 \label{NC1a}
        $$
        $$
        f_x(x_{\max}, y_{\max}) = 0, \quad f_{xx}(x_{\max}, y_{\max}) < 0 \label{NC1b}
        $$
        That is, $f(x, y)$ has a non-degenerate minimum at $x_{\min}$ and a non-degenerate maximum at $x_{\max}$. These correspond to critical points of the $f$-nullcline in the $x$-direction.
    \item \textbf{(NC2)} \textit{Tangential intersection of nullclines}:
    There exists a $C^{k}$ (for some \( k \geq 2 \)) function $g(x,y)$ such that, evaluated at the same $y = y_{\min}$ and $y = y_{\max}$, the function $g(x) := g(x, y)$ satisfies the following at the critical points $x_{\min}$ and $x_{\max}$:
    Matching function values (nullclines intersect):
    $$
    g(x_{\min}, y_{\min}) = f(x_{\min}, y_{\min}), \quad g(x_{\max}, y_{\max}) = f(x_{\max}, y_{\max}) \label{NC2a}
    $$
    Matching first derivatives (nullclines tangent):
    $$
    g_x(x_{\min}, y_{\min}) = f_x(x_{\min}, y_{\min}) = 0, \quad g_x(x_{\max}, y_{\max}) = f_x(x_{\max}, y_{\max}) = 0 \label{NC2b}
    $$
    This condition ensures that the $g$-nullcline intersects and is tangent to the $f$-nullcline at the extrema of $f$, i.e., at $x_{\min}$ and $x_{\max}$.
\end{itemize}

\paragraph{Summary.} The necessary conditions for the existence of a SNIC$^2$ are:
$$
\begin{aligned}
\textbf{(NC1)} &\quad \exists\, x_{\min}, x_{\max} \in \mathbb{R} \text{ such that:} \\
&\quad f_x(x_{\min}, y_{\min}) = 0,\quad f_{xx}(x_{\min}, y_{\min}) > 0, \\&\quad f_x(x_{\max}, y_{\max}) = 0,\quad f_{xx}(x_{\max}, y_{\max}) < 0; \\
\\
\textbf{(NC2)} &\quad g(x_{\min}, y_{\min}) = f(x_{\min}, y_{\min}),\quad g_x(x_{\min}, y_{\min}) = 0, \\
&\quad g(x_{\max}, y_{\max}) = f(x_{\max}, y_{\max}),\quad g_x(x_{\max}, y_{\max}) = 0.
\end{aligned}
$$
To complete the characterization of the SNIC$^2$ phenomenon, we could also specify the Jacobian-based conditions that ensure the existence of non-hyperbolic saddles at the points where the nullclines intersect tangentially, as described in \textbf{NC1} and \textbf{NC2}.

\paragraph{Jacobian conditions for non-hyperbolic saddles.} Let the Jacobian matrix of the system be:
\[
J(x, y) = 
\begin{pmatrix}
f_x(x, y) & f_y(x, y) \\
g_x(x, y) & g_y(x, y)
\end{pmatrix}.
\]
Then at both points \( (x_{\min}, y_{\min}) \) and \( (x_{\max}, y_{\max}) \), the following must hold:

\begin{itemize}
    \item \textbf{(J1)} Fixed point condition:
    \[
    f(x^*, y^*) = 0, \quad g(x^*, y^*) = 0, \quad \text{for } x^* \in \{x_{\min}, x_{\max}\}, y^* \in \{y_{\min}, y_{\max}\}
    \]
    
    \item \textbf{(J2)} Zero determinant (non-hyperbolicity):
    \[
    \det J(x^*, y^*) = f_x(x^*, y^*) g_y(x^*, y^*) - f_y(x^*, y^*) g_x(x^*, y^*) = 0.
    \]
    
    \item \textbf{(J3)} Non-zero trace (saddle-like character):
    \[
    \operatorname{tr} J(x^*, y^*) = f_x(x^*, y^*) + g_y(x^*, y^*) \neq 0.
    \]
\end{itemize}
Together, these conditions ensure the simultaneous occurrence of two saddle-node-on-invariant-circle bifurcations, giving rise to two non-hyperbolic saddle (saddle-node) equilibria and defining a SNIC$^{2}$ organizing center.

To illustrate the idea of SNIC$^2$ as an organizing center, assume the system of equations \ref{gen:eq1}-\ref{gen:eq2} is based on Wilson-Cowan model that describes the temporal dynamics arising from the interactions between populations of simplified excitatory and inhibitory neurons (see \hyperref[methodss]{Methods}). The Wilson-Cowan formalism and its extensions have become common frameworks for studying up and down state transitions \cite{jercog_up-down_2017, cakan_spatiotemporal_2022, ghorbani_nonlinear-dynamics_2012, levenstein_nrem_2019}. Wilson and Cowan demonstrated that their model is capable of supporting up to five equilibrium points, highlighting a rich repertoire of dynamical regimes accessible in the framework \cite{wilson_excitatory_1972}. 

Within the Wilson--Cowan framework, let $x$ and $y$ denote the activities of the excitatory and inhibitory populations, respectively, with the population interactions mediated by sigmoidal coupling functions. Sigmoidal response functions provide a biologically grounded representation of neuronal population activity, incorporating threshold-dependent activation, saturation at high input levels, and gradual recruitment of neurons \cite{marreiros_population_2008}. Phase space analysis of the system of equations \ref{gen:eq1}-\ref{gen:eq2} demonstrates that the excitatory nullcline is non-monotonic, exhibiting a maximum and a minimum, while inhibitory nullcline typically remains monotonic and has a sigmoidal profile \cite{wilson_excitatory_1972}. These nullcline geometries allow for the existence of a parameter regime in which the system exhibits a SNIC$^2$ organizing center (\autoref{fig:1}\textbf{B}(left)).
 
 We note that in this scenario the two saddle-node equilibria coexist with an additional equilibrium point located between them in the phase space (\autoref{fig:1}\textbf{B}(left)). Importantly, this intermediate equilibrium is unstable and therefore does not introduce an additional attractor. As a result, the global dynamics remains organized by the two saddle nodes and the heteroclinic connections between them, allowing the SNIC$^2$ structure to emerge. The system enters a bistable regime when both saddle nodes simultaneously split into a saddle and a stable node (\autoref{fig:1}\textbf{A} green region and \autoref{fig:1}\textbf{B}(right)). An example of such dynamics could also be found in a Wilson-Cowan-like model with adaptation used to simulate up and down state transitions during NREM sleep \cite{levenstein_nrem_2019}. The authors argue that experimentally observed transitions between up and down states do not necessarily require bistability. Instead, such transitions can emerge from excitable dynamics, where either the up state or the down state is the only stable equilibrium, while the system also has a saddle and an unstable equilibrium that shape the system's response to perturbations. We show that starting from the SNIC$^2$ point, if one saddle node is destroyed while the other is split into a saddle and a node, the system transitions into an excitable up (\autoref{fig:1}\textbf{A}, blue region) or excitable down state (\autoref{fig:1}\textbf{A}, purple region), where suprathreshold perturbations can generate a large excursion through a phase space prior to convergence back to the stable equilibrium. Another scenario is when both saddle nodes are destroyed simultaneously, as a result the system transitions into a mode with limit cycle solutions (\autoref{fig:1}\textbf{A}, orange region). Taken together, the above demonstrates that SNIC$^2$ provides a dynamical blueprint through which small changes in system parameters can reorganize network activity between qualitatively distinct regimes.

\subsection*{Examples of SNIC$^2$ Organizing Center in Neuronal Population Models}
Having introduced the general notion of SNIC$^2$ organizing center, together with its associated dynamics and necessary conditions, we now examine explicit examples of the phenomenological mean-field models in which this structure occurs. Using bifurcation and phase-plane analysis, we show that SNIC$^2$ is not model-specific, but arises across a broader class of systems. We further show that changes in excitation-inhibition balance govern transitions between dynamical regimes by reshaping the nullcline geometry and moving the system across bifurcation boundaries. This establishes excitation-inhibition balance as both a mechanistic determinant of the dynamics and a physiologically meaningful control parameter. 

\subsubsection*{Organization of activity regimes by excitatory drive and inhibitory coupling in the Wilson–Cowan model}\label{wc}

Having established the geometric conditions for SNIC$^2$ using the Wilson-Cowan model as an illustrative example, we next perform a two-parameter bifurcation analysis, using the external excitatory input and inhibitory connection strength as continuation parameters. This reveals how variations in excitatory drive and inhibitory feedback can organize transitions between distinct activity regimes. A detailed description of the model with the parameters used for the current piece of analysis is provided in \hyperref[methodss]{Methods}. In order to identify a parametrization that gives rise to a SNIC$^2$ organizing center, we build on our previous work in which we introduced an extended Wilson–Cowan framework for modeling population dynamics in the posterodorsal medial amygdala. In that model, we demonstrated the existence of a SNICeroclinic bifurcation (see Fig. S2 in the Supplemental Material of \cite{nechyporenko_neuronal_2024}), namely a heteroclinic loop connecting a saddle equilibrium and a saddle–node equilibrium. Using the obtained parameters as a starting point, we decrease the parameter responsible for the excitatory connection strength and arrive at a parameter configuration in \autoref{tab:wc_par} that gives rise to the presence of the SNIC$^2$ organizing center (\autoref{fig:2}). 

The bifurcation diagram demonstrates how the behavior of the system changes with varying inhibitory connection strength ($c_{IE}$) and the external (excitatory) input ($K_p$), depicting two sets of saddle node curves that divide the bifurcation diagram into four regions with distinct qualitative dynamics. The regime in which both sets of saddle equilibria coexist with stable nodes gives rise to bistability (\autoref{fig:2} region \textbf{I}). Depending on initial conditions the system can converge either to the up or down state, and the transition between these states can occur due to noise and/or slow modulation, such as adaptation, see \autoref{fig:1} \textbf{A}(green subregion) for an example time series simulation. Decreasing inhibitory connection strength drives the system into the canonical bistable regime where the basins of attraction of two stable equilibria are separated by the stable manifold of the saddle point (\autoref{fig:2} region \textbf{Ia}). To the contrary, increasing inhibition strength in the model triggers a saddle node bifurcation that eliminates the up state, leaving only the excitable down state (\autoref{fig:2} region \textbf{II}). The system now has one unstable, one saddle, and one stable equilibrium, supporting a down state while allowing for a perturbation exceeding the threshold to induce a large excursion through a phase space before the trajectory returns to the stable equilibrium. Further increasing inhibitory connection strength or reducing external input shifts the system to the canonical down state with a single stable (low-activity) equilibrium point.(\autoref{fig:2} region \textbf{IIa}). In this regime, any perturbation will converge monotonically toward the stable equilibrium and the long non-monotonic excursions associated with the nearby saddle structure are no longer present. 

Alternatively, an increase of the excitatory input drives the system from down state regime into oscillatory mode induced via saddle node on invariant circle (SNIC) bifurcation (\autoref{fig:2} region \textbf{III}). This transition occurs as excitation increasingly outweighs inhibition, destabilizing the quiescent state and giving rise to self-sustained oscillations. When the balance in the system shifts more towards excitation, by decreasing inhibitory connection strength, the system goes through another SNIC bifurcation leading to the annihilation of oscillations and emergence of the excitable up state (\autoref{fig:2} region \textbf{IV}). In this parameter region the system has one unstable, one saddle, and one stable equilibrium point, supporting high-activity regime while allowing for a transient excursion to a lower-activity regime following a sufficient perturbation. Upon further decrease in the inhibitory connection strength the system transitions to the canonical up (high-activity) state with only one stable equilibrium (\autoref{fig:2} region \textbf{IVa}). The disappearance of the saddle removes the phase-space structure responsible for the long transient excursions, leaving trajectories to converge monotonically to the stable equilibrium. The four dynamic regimes depicted in \autoref{fig:2}(\textbf{I}-\textbf{IV}) coalesce at the SNIC$^2$ point (\autoref{fig:2}  phase plane at the star point). At this organizing center the system spends infinite time at each saddle node, however, small level of noise can induce transitions, driving the system to switch sequentially between up and down states. (see \autoref{fig:1} \textbf{A}(pink subregion) for an example of time series simulation). 

Together, these results demonstrate that the underlying phase-space geometry is shaped by the relative strength of excitation and inhibition. Increasing inhibition stabilizes low-activity states and reduces the likelihood of transitions to high-activity regimes following perturbations, whereas increasing excitation promotes persistent high-activity states. The SNIC$^2$ point serves as a codimension-two organizing center, structuring the emergence of these dynamical regimes and the transitions between them.

\begin{figure*}[!h] 
    \centering
    \includegraphics[width=0.99\textwidth]{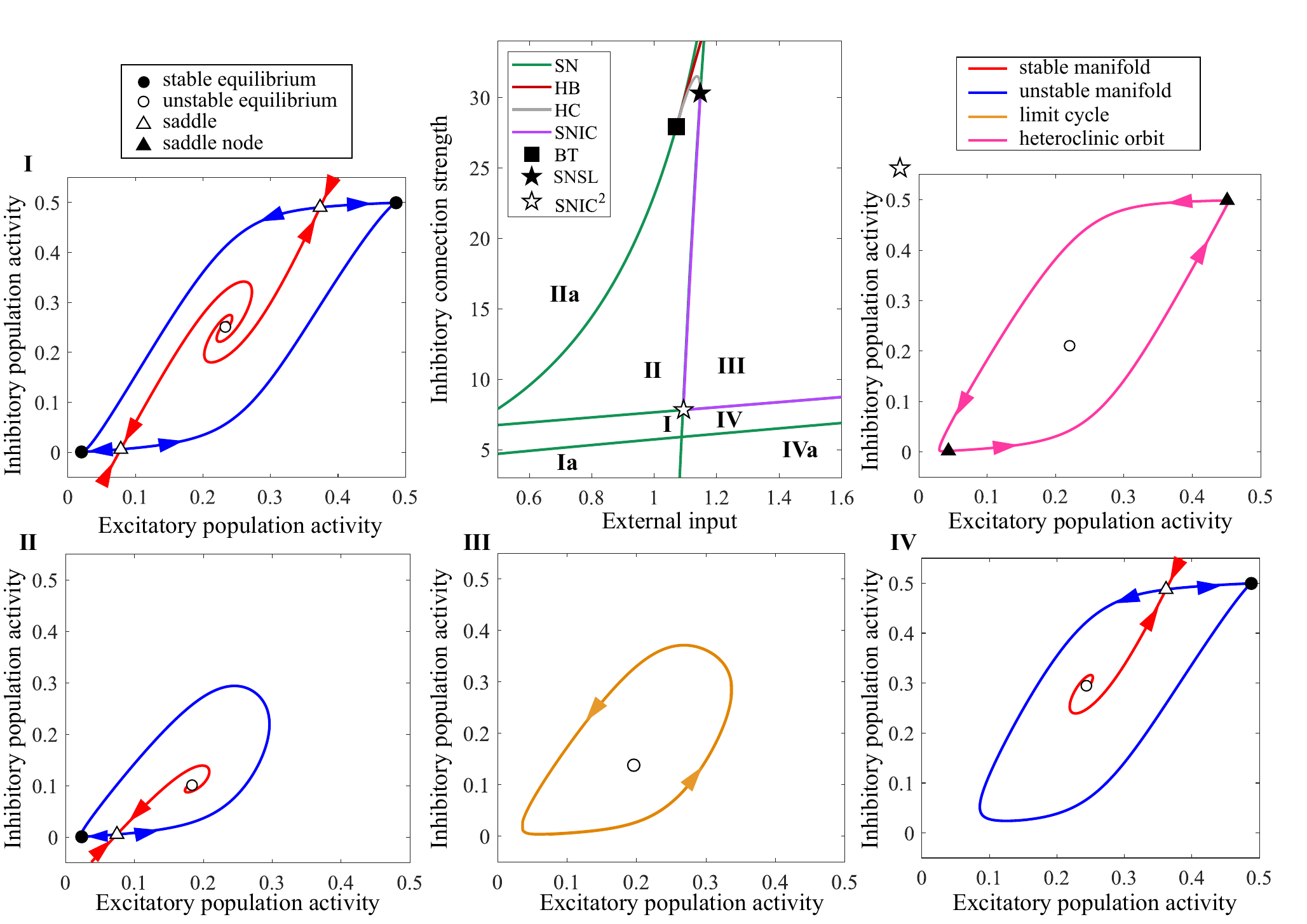}  
        \caption{Bifurcation diagram of the Wilson-Cowan model showing the dynamical regimes as a function of external excitatory input and inhibitory coupling strength, and revealing the SNIC$^2$ blueprint underlying their transitions. The bifurcation diagram shows arrangement of the saddle node (SN), Hopf (HB), homoclinic (HC) and saddle node on invariant circle (SNIC) curves. The points BT and SNSL correspond to the Bogdanov-Takens and saddle node separatrix loop bifurcations, respectively. The phase portraits demonstrates the dynamics in the regions surrounding SNIC$^2$ organizing center. The dynamics in the regions \textbf{Ia}, \textbf{IIa},\textbf{IVa} is excluded as there we find canonical bistability, down-only and up-only states, respectively.}
    \label{fig:2} 
\end{figure*}

\subsubsection*{Organization of activity regimes by excitatory drive and mean synaptic strength in the Tsodyks–Markram model}

We then consider Tsodyks-Markram model of synaptic plasticity \cite{tsodyks_neural_1997, tsodyks_neural_1998} to investigate the universality of the bifurcation structure we have revealed using the Wilson-Cowan model. While the original version of the Tsodyks-Markram model includes average synaptic activity, modulated by depression and facilitation variables, we use the reduced version of the model proposed by Holcman and Tsodyks \cite{holcman_emergence_2006}. In this version of the Tsodyks-Markram model the facilitation variable is considered to operate on much slower time-scale, which is consistent with known properties of the synapses in the prefrontal cortex \cite{wang_heterogeneity_2006}. Consequently, facilitation is represented by a parameter in the model. The detailed model description and corresponding model parameters are given in \hyperref[methodss]{Methods}. 

We demonstrate the emergence of a SNIC$^2$ structure by performing two-parameter bifurcation analysis using the external (excitatory) drive ($I$) and mean synaptic strength, ($\omega$) as control parameters, (\autoref{fig:3}). Biologically, the mean synaptic strength determines how strongly recurrent synaptic activity contributes to the amplification and maintenance of collective network activity. The resulting bifurcation diagram closely resembles the arrangement of dynamical regimes surrounding the SNIC$^2$ point in the Wilson-Cowan model (\autoref{fig:2}), demonstrating that different model-specific parameters can nevertheless organize similar transitions between dynamical regimes in both models.

The system can exhibit bistability with two stable, one unstable, and two saddle equilibria (\autoref{fig:3}(\textbf{I})). As the average synaptic strength decreases, the system transitions to an excitable low-activity (down) state through a saddle-node bifurcation that destroys the up state (\autoref{fig:3}(\textbf{II})). Similar to increasing inhibitory coupling in the Wilson-Cowan model, this parameter change effectively shifts the balance between excitation and inhibition towards inhibition, favoring low-activity dynamics. Although there is only one stable attractor in this regime, the presence of an unstable and a saddle equilibria allows transient excursions towards the up state. A further decrease in synaptic strength destroys the hyperbolic saddle and unstable equilibrium through a saddle node bifurcation and drives the system into the canonical down-state regime, where only a single stable equilibrium remains (\autoref{fig:3}(\textbf{IIa})). On the other hand, increasing the external input leads to oscillations induced via a SNIC bifurcation (\autoref{fig:3}(\textbf{III})). Further increase in the external drive results in the appearance of another SNIC bifurcation and a transition into the  excitable up state corresponding to an elevated mean population activity and low depression (\autoref{fig:3}(\textbf{IV})). Increasing even further the external input drives the system through the saddle node bifurcation, leading to a single stable equilibrium (\autoref{fig:3}(\textbf{IVa}). The SNIC$^2$ point (see the star in the phase plane in \autoref{fig:3}) marks the codimension-two organizing center at which the bifurcation boundaries separating these qualitatively different activity regimes meet.

Despite capturing different activity-dependent aspects of neural function, the Tsodyks-Markram and Wilson-Cowan models exhibit similar dynamical blueprint. In both modeling frameworks, transitions between distinct activity regimes are controlled by an effective balance between activity-promoting and activity-suppressing drives. While this balance is represented explicitly through excitatory and inhibitory populations in the Wilson-Cowan model, in the Tsodyks-Markram model it emerges through the interaction between neuronal activity and activity-dependent synaptic depression. As this balance shifts, both models display analogous dynamical regimes, including bistability, up and down states, and limit cycle solutions. In each case, the SNIC$^2$ point acts as a codimension-two organizing center for the bifurcation boundaries separating activity regimes, revealing a shared dynamical structure across different biological formulations.

\begin{figure*}[!h] 
    \centering
\includegraphics[width=0.99\textwidth]{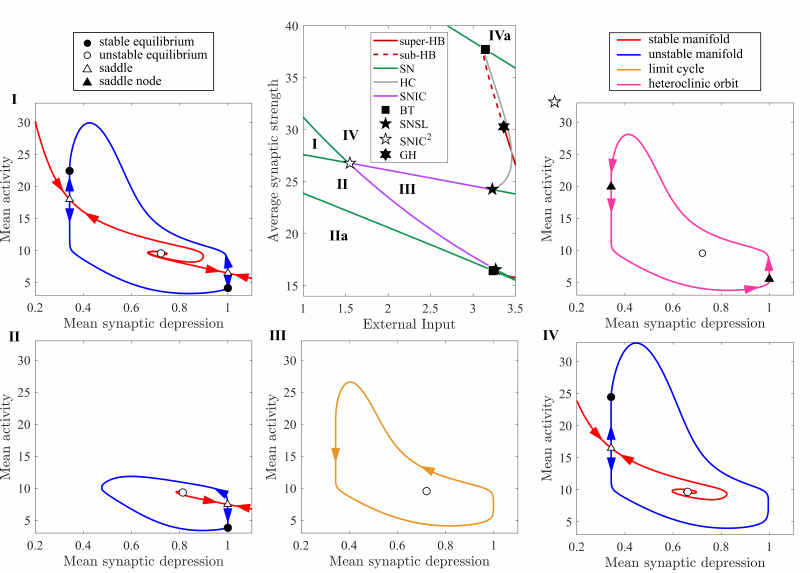}  
    \caption{Visualization of the dynamics of the Tsodyks-Markram model around the SNIC$^2$ blueprint identified in the two-parameter bifurcation diagram spanned by the external excitatory input and the average inhibitory synaptic strength. The bifurcation diagram shows arrangement of the saddle node (SN), super- and subcritical Hopf (HB), homoclinic (HC) and saddle node on invariant circle (SNIC) curves. The points BT, SNSL and GH correspond to the Bogdanov-Takens point, saddle node separatrix loop point and generalized Hopf, respectively. The phase portraits illustrate the dynamics of the regions surrounding SNIC$^2$ organizing center. The regions \textbf{IIa} and \textbf{IVa} are excluded as there we find a canonical down state and a canonical up state, respectively.}
    \label{fig:3} 
\end{figure*}

\subsubsection*{Astrocytes as a switchboard for neuronal population activity}

Traditionally regarded as passive support cells, astrocytes are now recognized as active participants in regulating neural network dynamics. For example, in the basolateral amygdala they have been shown to contribute to the generation and adaptation of fear-state-related neural representations \cite{bukalo_astrocytes_2026}, highlighting their role in shaping and reorganizing population-level activity patterns. Growing evidence indicates that neurons and astrocytes engage in bidirectional communication: neuronal activity triggers neurotransmitter release that evokes astrocytic Ca$^{2+}$ elevations~\cite{rupareliya_molecular_2023}, while astrocytes, in turn, modulate neuronal activity via gliotransmission~\cite{goenaga_calcium_2023}. From a dynamical systems perspective, astrocytes therefore provide an additional regulatory mechanism capable of modulating the balance between excitation and inhibition at the network level. Extending phenomenological mean-field models to include astrocytic interactions allows this regulation to be studied explicitly. In this setting, astrocytes introduce an additional dynamical degree of freedom that can reshape the phase-space structure and influence transitions between qualitative dynamical states. Furthermore, incorporating astrocytes enables us to investigate whether the geometric structures associated with a SNIC$^2$ organizing center can persist in a higher-dimensional setting.

 Previous work extended the Wilson-Cowan model with adaptation to include astrocyte interactions with excitatory and inhibitory populations \cite{blum_moyse_modelling_2022}. In that framework, oscillations arise through negative feedback between the excitatory population and a slow adaptation variable. However, the adaptation variable is coupled linearly to the excitatory population, preventing the formation of two tangential nullcline intersections. Consequently, the model lacks the geometric conditions required for a SNIC$^2$ organizing center. To investigate whether astrocyte-mediated regulation can support such structures, we instead extend the classical Wilson-Cowan model without adaptation \cite{wilson_excitatory_1972} by introducing an astrocytic population.

The astrocytic population ($A$) integrates input from both excitatory ($E$) and inhibitory ($I$) neuronal populations and feeds back onto each population, thereby modulating their activity (see \hyperref[methodss]{Methods}, \autoref{fig:astro_sup1}(\textbf{A})). When excitatory and inhibitory neurons fire, they release neurotransmitters such as glutamate and GABA into the synaptic cleft, where they bind to receptors on nearby astrocytic processes within the tripartite synapse. Through these receptors, neuronal activity is integrated within astrocytes involving intracellular Ca$^{2+}$ dynamics \cite{rupareliya_molecular_2023}. As a result, in response to calcium elevations, astrocytes can release gliotransmitters that modulate neuronal membrane excitability \cite{goenaga_calcium_2023}, thereby influencing the excitation-inhibition balance governing network dynamics. In the model, this regulation is represented phenomenologically through astrocyte-to-neuron interaction strengths that describe the net astrocytic influence on excitatory and inhibitory populations. While this formulation does not capture the full complexity of neuron-astrocyte signaling, it provides a simplified description of the bidirectional interactions through which astrocytes can modulate network activity, similar to the approach adopted in \cite{blum_moyse_modelling_2022}. The detailed model description can be found in \hyperref[methodss]{Methods} along with the model parameters given in \autoref{tab:wc_par_astro}.

 Using bifurcation analysis, we focus on a regime in which astrocytes exert a net excitatory influence on the network. As a baseline, we consider parameter values for which the neuronal subsystem, in the absence of astrocytes, exhibits only stable equilibrium dynamics because the inhibitory feedback is insufficient to generate oscillations (\autoref{fig:astro_sup1}(\textbf{B})). By considering a neuronal network regime that does not intrinsically generate rhythmic activity, we can more clearly assess how astrocytic feedback modifies the network dynamics, independent of oscillatory mechanisms already present in the neuronal subsystem.
 
 Next, by varying the astrocyte-to-excitatory ($c_{AE}$) and astrocyte-to-inhibitory ($c_{AI}$) interaction strengths, we recover the same bifurcation structure around the SNIC$^2$ point as is identified in the classical Wilson-Cowan model through changes in inhibitory strength and external (excitatory) drive (\autoref{fig:astro}). We start from the bistable regime, where the system is able to transition between up and down states subject to perturbations (\autoref{fig:astro} region \textbf{I}). In our three-dimensional system both saddles are characterized by a one-dimensional unstable manifold and a two-dimensional stable manifold (\autoref{table_eig1star}). Since the stable manifolds form two-dimensional surfaces in the full three-dimensional phase space (span by $E$, $I$, and $A$), their structure is obscured in two-dimensional projections involving E and A only. We therefore visualize the one-dimensional unstable manifolds instead, as they more clearly highlight the global phase space geometry. We note that, in contrast to the two-dimensional Wilson-Cowan and Tsodyks-Markram models, the unstable manifold of the saddle associated with the down-state stable equilibrium does not connect to the up-state stable equilibrium. Instead, it makes an excursion through the phase plane before converging to the down-state stable equilibrium. This behavior is likely possible due to the additional degree of freedom introduced by the astrocytic variable, which modifies the global phase-space geometry. The complete set of projections of the three-dimensional phase space is provided in \autoref{fig:astro_sup2} in \hyperref[SI]{Supporting Information}.

Decrease in the interaction strength from astrocytic population to the excitatory population drives the system through saddle node bifurcation destroying the up state, leaving just an excitable down state (\autoref{fig:astro} region \textbf{II}). A qualitatively similar transition is achieved by increasing inhibitory connection strength in Wilson-Cowan and decreasing synaptic strength in Tsodyks-Markram, with all three cases shifting the excitation-inhibition balance towards inhibition. Now, decrease in the interaction strength from astrocytic population to the inhibitory population, drives the system through SNIC bifurcation, inducing a limit cycle solutions (\autoref{fig:astro} region \textbf{III}). Further decrease in the interaction strength from astrocytic population to the inhibitory population or increase in the interaction strength from astrocytic population to the excitatory population forces the system through another SNIC bifurcation, destroying the oscillatory behavior and generating an excitable up state dynamics (\autoref{fig:astro} region \textbf{IV}).

Similarly to the two-dimensional models, the qualitative regimes described above coalesce at the SNIC$^2$ organizing center (see \autoref{fig:2} phase plane at the star point). The regions in the bifurcation diagram \textbf{Ia}, \textbf{IIa}, \textbf{IVa} identify the regimes of canonical bistability, down state and up state, respectively, as in the original Wilson-Cowan framework. At the SNIC$^2$ the saddle nodes possess a two-dimensional stable manifold and one-dimensional center manifold (\autoref{table_eig2star}). Despite the additional dimension, the SNIC$^2$ loop remains one-dimensional, with its organization governed by the center manifolds of the saddle nodes. 

We note that an investigation of another higher-dimensional system, namely the six-dimensional phenomenological mean-field Jansen-Rit model (\hyperref[SI]{Supporting Information}), reveals the persistence of a SNIC$^2$ organizing center. Similarly to Wilson-Cowan with astrocytes, the SNIC$^2$ loop we discover in the Jansen-Rit model is organized by a one-dimensional center manifold, with the stable manifolds of the saddle nodes involved being five-dimensional.  

In summary, our findings demonstrate that astrocyte-mediated feedback can regulate network dynamics in an analogous manner to changes in connection (synaptic) strength and constant external input, providing an additional mechanism through which neuronal networks can transition between qualitatively distinct activity states. Although astrocytes introduce an additional dynamical variable, the underlying SNIC$^2$ blueprint is preserved.

\begin{figure*}[h] 
    \centering
    \includegraphics[width=0.99\textwidth]{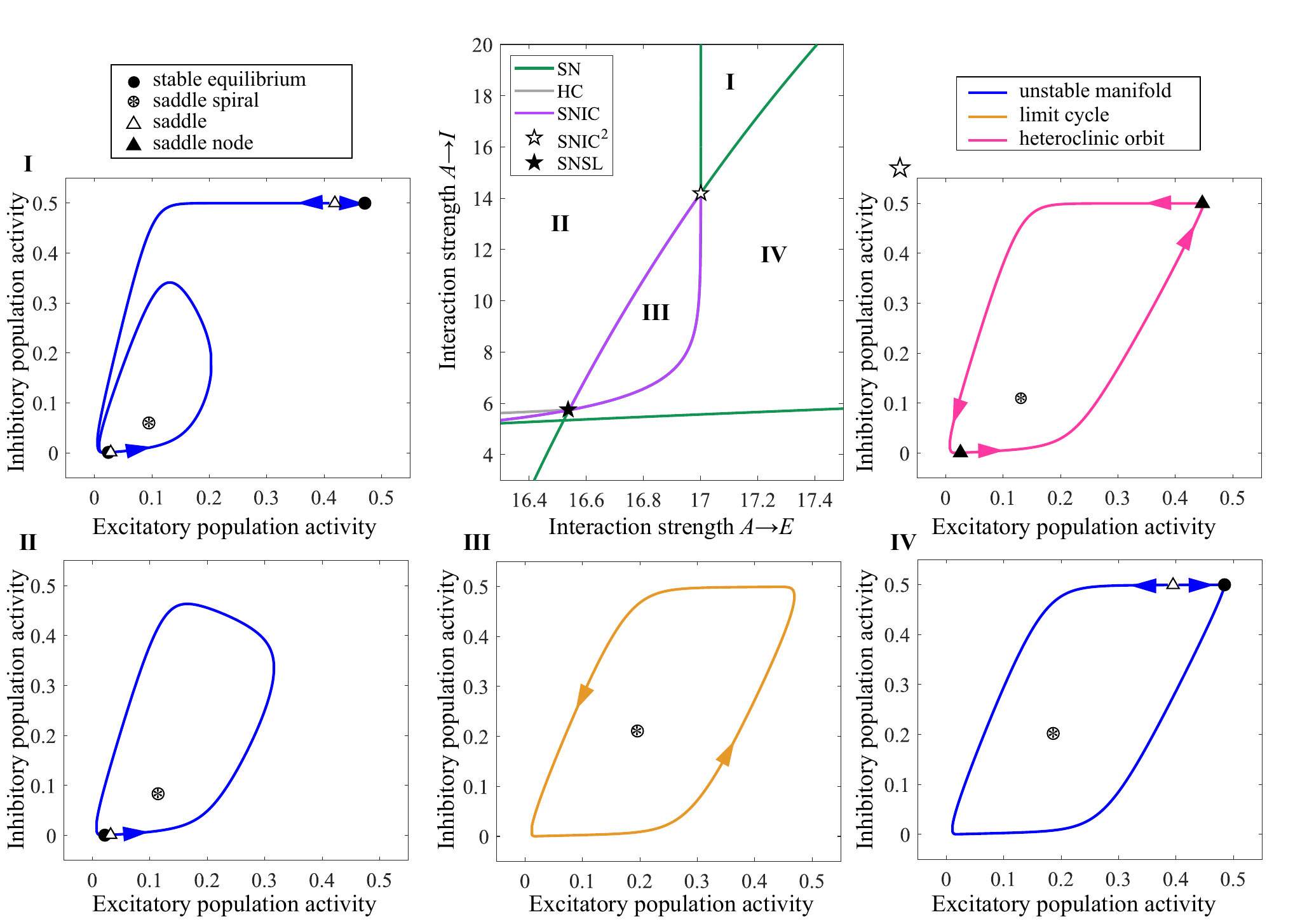}  
        \caption{Visualization of the astrocytes modulating the dynamics in the neuronal population, organized by the SNIC$^2$. Bifurcation diagram demonstrates how interaction strength from astrocytes to the neuronal populations ($c_{AE}$ for excitatory and $c_{AI}$ inhibitory neuronal populations) shapes the dynamics, showing an arrangement of the saddle node (SN) curves with saddle node on invariant circle (SNIC) and homoclinic (HC) curves. SNSL point identifies a saddle node separatrix loop.}
    \label{fig:astro} 
\end{figure*}

\subsection*{Universality of the phenomenon}

We have shown that excitation-inhibition balance provides a common mechanism across four phenomenological mean-field neuronal network models: the Tsodyks-Markram model, the Jansen-Rit model, the original Wilson-Cowan model, and its astrocyte-augmented extension. In each case, variations in this balance give rise to qualitatively similar transitions between activity regimes, revealing SNIC$^2$ as a shared dynamical blueprint for regime organization.
While the phenomenological mean-field models considered here employ sigmoidal nonlinearities, the SNIC$^2$ mechanism is not tied to a particular functional form. Rather, it emerges whenever the geometric conditions \hyperref[NC1a]{\textbf{NC1-NC2}} are satisfied, allowing the phenomenon to arise across diverse modeling frameworks. For example, we have also identified a SNIC$^2$ point in the Morris-Lecar model \cite{morris_voltage_1981}, a two-dimensional conductance-based model of excitable-cell dynamics that couples membrane-potential evolution to ion-channel gating kinetics and is widely used to describe spiking behavior in neurons and muscle fibers. This model employs hyperbolic tangent functions that are akin to the sigmoid given by exponential functions. Here, we obtain a two-parameter bifurcation diagram that is similar to that of the phenomenological mean-field network models by varying the external (excitatory) drive to the cell and the maximal potassium conductance, which effectively modulates the strength of the potassium (inhibitory) current (\hyperref[SI]{Supporting Information}). This finding demonstrates that the SNIC$^2$ organizing structure is not restricted to phenomenological mean-field models, but may also arise in lower-dimensional biophysical models of neuronal excitability, opening possibilities beyond neurodynamics. Across all five models, the SNIC$^2$ emerges under variations of parameters that regulate the balance between excitatory and inhibitory drivers in the system. These drivers may act through external input, astrocytic modulation, connection or synaptic strengths, or ion-channel conductances, yet they produce the same qualitative bifurcation structure.

\section*{Discussion}

The key finding of this work is the identification of a heteroclinic loop connecting non-hyperbolic saddles, termed SNIC$^2$, as a previously unrecognized blueprint for neuronal dynamics that organizes transitions between different dynamical regimes through changes in the excitation-inhibition balance. By deriving necessary conditions for its existence and demonstrating its emergence across multiple neuronal modeling frameworks, we show that distinct dynamical regimes can be linked through a common organizing structure, thereby advancing our understanding of neuronal state organization and suggesting broader relevance to nonlinear dynamical systems beyond neuroscience.

Our organizing center builds on existing theory of non-chaotic metastability by unifying dynamical scenarios that have previously been studied majorly in isolation. Specifically, \cite{rossi_dynamical_2025} contrasts noise-driven bistability between two stable equilibria with sequential switching dynamics organized by a stable heteroclinic cycle connecting hyperbolic saddles. In comparison, the SNIC$^2$ structure identified here provides a geometric blueprint that incorporates both bistability and sequential transitions organized by a non-hyperbolic heteroclinic loop. Unlike classical codimension-one (Hopf, SNIC, and homoclinic) and codimension-two (saddle-node loop, Bogdanov-Takens, generalized Hopf, and cusp) bifurcations \cite{saggio_new_2026}, which are generally used to explain individual aspects of neuronal dynamics, SNIC$^2$ is a global bifurcation that unifies transitions between bistability, oscillations, up and down states within a single geometric structure and additionally can give rise to excitability, where perturbations can induce large transient excursions before relaxation to a stable equilibrium. In this sense, SNIC$^2$ serves as a central organizing center from which multiple dynamical regimes, relevant to neuronal network behavior, emerge through relatively small parameter variations.

Studies of heteroclinic phenomena in the planar (two-dimensional) case have primarily focused on loops involving hyperbolic saddle equilibria \cite{homburg_chapter_2010}. The SNIC$^2$ mechanism constitutes a further increase in degeneracy, arising from the interaction of two saddle–node on invariant circle mechanisms. To our knowledge, such structures remain largely unexplored in the applied dynamical systems literature. More broadly, our results establish a connection between the theory of non-hyperbolic heteroclinic structures and physiologically relevant switching dynamics. In planar systems, SNIC$^2$ emerges across a range of neuronal models, including the phenomenological Wilson-Cowan and reduced Tsodyks-Markram mean-field models \cite{wilson_excitatory_1972, tsodyks_neural_1997, holcman_emergence_2006} as well as the biophysical Morris-Lecar neuron model \cite{morris_voltage_1981}. This recurrence across models suggests that SNIC$^2$ is a robust dynamical mechanism rather than a model-dependent feature, while the necessary conditions established here point to its potential occurrence in nonlinear systems far beyond the neuronal context. Many biological models, including genetic toggle-switch circuits \cite{gardner_construction_2000}, GTPase activation networks \cite{jilkine_comparison_2011}, and predator-prey systems with an Allee effect \cite{arancibia-ibarra_hollingtanner_2019}, exhibit threshold-like interactions that give rise to nonlinear phase-space geometries that can potentially support SNIC$^2$.

From a dynamical systems perspective, the SNIC$^2$ phenomenon opens new directions for studying the interplay between excitability, multistability, and oscillatory behavior beyond planar systems. While the higher-dimensional models (extended Wilson-Cowan \cite{wilson_excitatory_1972} with astrocytes and Jansen-Rit \cite{jansen_electroencephalogram_1995}) considered here are organized by one-dimensional center manifolds and therefore reproduce the planar SNIC$^2$ configuration, other possibilities may arise when saddle nodes possess unstable non-central eigendirections and/or higher codimension organizing centers where one or both of the heteroclinic connections occur in non-central directions. In \cite{nechyporenko_novel_2026} we describe the standard (codimension-two) and non-central (codimension-three) SNICeroclinic bifurcations involving a saddle and a saddle node, and give various examples, including a GTPase signaling model, where the resulting periodic orbit is unstable. In such a case, the heteroclinic structures play important roles in organizing basin boundaries. In higher-dimensional systems, there is also a possibility of coexistence of multiple heteroclinic connections between saddle-node equilibria, leading to substantially richer global bifurcation structures. These observations suggest that SNIC$^2$ may represent only the simplest case of a broader class of non-hyperbolic heteroclinic structures. 

An important question is how such organizing centers relate to the physiological mechanisms. Previous studies have identified noise and adaptation as key contributors to transitions between up and down dynamics \cite{jercog_up-down_2017, levenstein_nrem_2019, holcman_emergence_2006}. Our work instead investigates how modulation of excitation-inhibition balance can configure the dynamical landscape that gives rise to qualitatively distinct regimes. Excitation-inhibition balance is a central physiological control mechanism in neural circuits because it determines whether recurrent activity is amplified, stabilized, suppressed, or allowed to propagate through the network \cite{van_vreeswijk_chaos_1996, shu_turning_2003}. In this sense, shifts in excitation-inhibition balance do not merely perturb the dynamics; they can govern the transition between qualitatively distinct regimes, such as quiescence, sustained activity, bistability and oscillations. By contrast, adaptation introduces slow activity-dependent negative feedback, for example through spike-frequency adaptation, synaptic depression, or other fatigue-like mechanisms \cite{holcman_emergence_2006, jercog_up-down_2017, brunel_dynamics_2000, sanchez-vives_cellular_2000}. These processes can gradually reduce network excitability following sustained activity, thereby promoting transitions out of active states or shaping the timing of recurrent switching. In contrast, noise need not modify the underlying deterministic bifurcation structure. Instead, it can drive stochastic excursions around it, trigger transitions between coexisting attractors, or expose latent instability when the system operates close to a bifurcation boundary.

Thus, while adaptation and noise are modulatory mechanisms that can shape the timing and expression of dynamical regimes, excitation-inhibition balance acts as a physiologically meaningful organizing parameter of the underlying network dynamics. This is particularly relevant in neural systems because many brain-state transitions can be understood as changes in the relative dominance of excitation and inhibition. Consistent with this view, excitation-inhibition balance has been widely linked to cortical stability, irregular activity, oscillatory transitions, critical-like dynamics, and brain-state regulation \cite{van_vreeswijk_chaos_1996, shu_turning_2003, liang_excitationinhibition_2025, haider_neocortical_2006, dehghani_dynamic_2016, deco_how_2014, markicevic_cortical_2020}, whereas noise and adaptation are thought to trigger, shape, or facilitate transitions once the network is already operating near a bistable or bifurcation regime. Therefore, adaptation and noise can influence when and how transitions occur, but excitation-inhibition balance determines the physiological operating regime in which those transitions become possible.

From a real-world application perspective, the importance of our findings spans several domains in neuroscience, medicine, and artificial intelligence (AI). Dysregulation of excitation-inhibition balance has been implicated in depression \cite{duman_altered_2019}, epilepsy \cite{van_van_hugte_excitatoryinhibitory_2023} and neurodevelopmental disorders \cite{markicevic_cortical_2020} and is associated with abnormal transitions between brain network activity states. Defining a dynamical blueprint that describes how excitation-inhibition balance shapes transitions between activity states could provide a mechanistic framework for understanding how such transitions become disrupted in disease and may help identify dynamical biomarkers or intervention targets. For example, astrocytes regulate excitation-inhibition balance through neurotransmitter uptake, potassium buffering, metabolic coupling, and gliotransmission \cite{stogsdill_astrocytes_2023}, and the SNIC$^2$ organizing center may provide a principled way to investigate how astrocyte-mediated modulation of network excitability can be leveraged to restore healthy dynamical regimes in pathological conditions. Additionally, understanding the dynamical principles organizing transitions in neuronal networks can inform how deep brain stimulation (DBS), transcranial magnetic stimulation (TMS), or closed-loop neurofeedback protocols are applied, for example in conditions like Parkinson’s or treatment-resistant depression. Health professionals could tailor stimulation parameters to push the system toward desired activity states enabled through modeling patient-specific dynamics and identifying how close a network is to a bifurcation point. Techniques such as bifurcation analysis and mean-field modeling offer a theoretical framework to decode brain-wide dynamics from macro-scale recordings \cite{dunstan_neural_2025,bick_understanding_2020,coombes_next_2023,frascoli_metabifurcation_2011}, thereby improving diagnostics or real-time brain monitoring (e.g., in ICU or surgery). This approach enables the connection of microscopic mechanisms (synapses) to macroscopic outcomes (cognitive state transitions). This in turn, helps unify models across scales, which is a major challenge in systems neuroscience. 

Last but not least, insights into the mechanisms driving state transitions in neuronal network dynamics based on global dynamic structures (rather than local rules) may contribute to the development of smarter and more brain-like AI. In particular, it could enable artificial neural systems to autonomously switch between exploration and exploitation, or between learning and inference, in a context-dependent and energy-efficient manner. Embedding similar principles in artificial systems could lead to adaptive computation, where the network dynamically enters different operating modes in response to task demands or sensory inputs, much like brain networks. SNIC$^2$ blueprint may therefore provide a framework for designing artificial neural systems capable of robust and flexible transitions between computational states arising from intrinsic network dynamics, rather than requiring explicit supervisory control, hard-coded switching rules, or externally imposed changes in system parameters.

In summary, we introduce a unifying organizing principle that governs how brain networks dynamically transition between different modes of activity. This could find applications in:
\begin{itemize}
    \item Diagnosing and treating brain disorders.
    \item Developing more effective brain-computer interfaces and neuromodulation tools.
    \item Enhancing our fundamental understanding of cognition and consciousness.
    \item Designing smarter and more brain-like AI.
\end{itemize}

\section*{Methods}\label{methodss}

\subsection*{Wilson-Cowan model: firing-rate model of excitatory-inhibitory population dynamics}
The Wilson-Cowan model \cite{wilson_excitatory_1972} provides a coarse-grained description of the network dynamics of interacting excitatory and inhibitory neuronal populations:
\begin{align}
    \frac{dE}{dt} &= -E+(1-E)\sigma(a_E,c_{EE}E-c_{IE}I+K_p\alpha,\theta_E) \\
    \frac{dI}{dt} &= -I+(1-I)\sigma(a_I,c_{EI}E-c_{II}I+K_p(1-\alpha),\theta_I),
\end{align}
where $E$ and $I$ represent the activity of the excitatory and inhibitory populations, respectively. The parameters $c_{ij}$ correspond to the strength of connection from the population $i$ to the population $j$ for $i,j \in \{E,I\}$, while $K_p$ and $\alpha$ define the magnitude of the external input to the populations. The function $\sigma$ is an activation response function of the form: 
\begin{align}
     \sigma(a, x, \theta) = \dfrac{1}{1+\exp(-a(x-\theta))}  - \dfrac{1}{1+\exp(a\theta)}, \label{sigmaaaa}
\end{align}
where $x$ indicates the input to a given population and parameters $a$ and $\theta$ define the value of maximum slope and half-maximum firing threshold, respectively.

\subsection*{Tsodyks-Markram: model of short-term plasticity}
We consider Tsodyks–Markram model, which describes short-term synaptic plasticity by capturing dynamic changes in synaptic efficacy through activity-dependent facilitation and depression, governed by biologically motivated differential equations \cite{tsodyks_neural_1997,tsodyks_neural_1998}. For our analysis, we use a smooth version of the model, introduced in \cite{holcman_emergence_2006} that considers a homogeneous population of neurons, characterized by a mean firing rate activity $V$, The second variable in the model accounts for synaptic depression, described by the mean rate $m$ across the network. The system is given by:
\begin{align}
    \frac{dV}{dt} &= (-V + mU\omega R_1(V)+\sqrt{\tau}+I)/\tau \\
    \frac{dm}{dt} &= \frac{1-m}{\tau_r}-mU R_2(V), 
\end{align}
where $V$ is the mean activity and $m$ is mean depression rate. In this variant of the model, facilitation is assumed to decay much more slowly than depression, a property observed at synapses in the prefrontal cortex \cite{wang_heterogeneity_2006} for example. Consequently, facilitation is represented as a constant parameter $U$ rather than as a dynamic variable, as in the original formulation \cite{tsodyks_neural_1997,tsodyks_neural_1998}. The parameter $\omega$ accounts for the average synaptic strength in the network, $\tau_r$ is the recovery time constant of synaptic depression, $\tau$ is the recovery time constant of the population, and $I$ is the external input. The function $R(V)$ represents an average firing rate and in \cite{holcman_emergence_2006} and \cite{mongillo_synaptic_2008} $R(V)$ is given by the threshold-linear function (i.e Heaviside function). For the purposes of numerical continuation analysis employing AUTO \cite{doedel_auto-07p_2007} we substitute the non-smooth threshold-function by a smooth sigmoidal function as given below. 

The function $R(V)$ in our case is the following: 
    \begin{equation}
        R(V)=R(v_{\max},V_0,V,r)=\frac{v_{\max}}{1 + \exp(r (V_0 - V))},
    \end{equation}
where $r$ is the reciprocal of the activation slope, $V_0$ is half-maximum activation and $v_{\text{max}}$ is the maximum threshold.
From a physiological point of view, presynaptic depression and postsynaptic activity are distinct biological processes, and therefore do not necessarily depend on the firing rate $V$ in the same way. Accordingly, the presynaptic firing that drives synaptic depression in the version of the model we consider here assumes slightly different parametrization of the input–output relationship than the one governing the overall population activity. The model parameters are given in \autoref{tab:tm_par}.

\subsection*{Extension of the Wilson-Cowan model incorporating astrocyte dynamics}
In order to study the effects of astrocytic modulation on neuronal networks, we extend Wilson-Cowan model \cite{wilson_excitatory_1972} by adding a third variable $A$, which describes the mean activity of the astrocytic population: 
 \begin{align}
    \frac{dE}{dt} &= -E+(1-E)\sigma(a_E,c_{EE}E-c_{IE}I+c_{AE}A+\alpha K_{p},\theta_E) \\
    \frac{dI}{dt} &= -I+(1-I)\sigma(a_I,c_{EI}E-c_{II}I+K_{p}(1-\alpha),\theta_I),\\
    \frac{dA}{dt} &= -A+(1-A)\sigma(a_A,c_{EA}E,\theta_A),
\end{align}
 where the excitatory neuronal population $E$ and inhibitory neuronal population $I$ provide input to the astrocytic population in proportion to the neuronal drive $c_{EA}$ and $c_{IA}$, respectively, representing the activity-dependent release of neurotransmitters that can activate astrocytic dynamics. In turn, the astrocytic population exerts a modulatory influence on the excitatory and inhibitory neuronal populations that is proportional to the strength of astrocytic input parameter $c_{AE}$ and $c_{AI}$, respectively, capturing gliotransmission-mediated feedback that regulates neuronal excitability. This bidirectional coupling allows the model to account for neuron-astrocytes population-level bi-directional interaction that shapes network dynamics. The sigmoid function $\sigma$ is given in \autoref{sigmaaaa}, the model parameters are given in \autoref{tab:wc_par_astro}. 
\subsection*{Numerical continuation and simulations}
The bifurcation analysis of the model systems was performed in AUTO-07p \cite{doedel_auto-07p_2007}, while the simulations are run in MATLAB. The code reproduce the analysis within the manuscript is in a github repository at \url{https://github.com/ktrnnchprnk/SNICsquared}.

\clearpage
\printbibliography
\subsection*{Acknowledgements:} KTA and PA gratefully acknowledge the financial support of the EPSRC via grant EP/T017856/1. We would like to acknowledge helpful comments and suggestions by Jan Sieber, Stephen Coombes and Kyle Wedgwood.

\fi
\clearpage
\setcounter{page}{1}
\renewcommand{\thepage}{S\arabic{page}}

\ifsupp

\maketitle  
\section*{Supporting Information (SI)}\label{SI}

\subsection*{Figures and  Tables}

\setcounter{figure}{0}
\renewcommand{\thefigure}{S\arabic{figure}}
\renewcommand{\thetable}{S\arabic{table}}

\begin{table}[!h]
    \centering
    \begin{tabular}{c|c|c}
   Parameter & Value & Description\\
    \hline
$a_e$ &  1.3 & maximum slope in the excitatory population\\
$a_i$ & 2 &   maximum slope in the inhibitory population\\
$\theta_e$& 4 &  half-maximum firing threshold in the excitatory population\\
$\theta_i$ & 3.7 &  half-maximum firing threshold in the inhibitory population \\
$c_{EE}$ & 18 &  self-excitation strength \\
$c_{EI}$ & 14 & excitatory connection strength  \\
$c_{II}$& 0 &self-inhibition \\
$\alpha$& 0.9 & proportion of the external input to the excitatory population \\
$K_p$ & [0,2] & excitatory (external) input  \\
$c_{IE}$& [0 40] &  inhibitory connection strength \\
    \end{tabular}
    \caption{Model parameters of Wilson-Cowan model.}
    \label{tab:wc_par}
\end{table}

\begin{table}[!h] 
    \centering
    \begin{tabular}{c|c|c}
   Parameter & Value & Description\\
    \hline
$U$ &  0.04 & facilitation parameter\\
$\tau$& 0.2 &  firing-rate recovery time constant\\
$\tau_r$ & 0.8 &  synaptic depression recovery time constant \\
$v_{\text{max}}$ & 60 & maximum threshold  \\
$V_{01}$& 15 &  firing-rate half-maximum activation \\
$V_{02}$& 10 & depression half-maximum activation \\
$r_{1}$& 0.3 & firing-rate reciprocal of the activation slope \\
$r_{2}$ & 3 & depression reciprocal of the activation slope\\
$I$& [0,3.5] & excitatory input \\
$\omega$& [10 40] & average synaptic strength \\
    \end{tabular}
    \caption{Model parameters of Tsodyks-Markram model.}
    \label{tab:tm_par}
\end{table}
\newpage
\begin{table}[!h]
    \centering
    \begin{tabular}{c|c|c}
   Parameter & Value & Description\\
    \hline
$a_e$ &  1.3 & maximum slope in the excitatory population\\
$a_i$ & 2 &   maximum slope in the inhibitory population\\
$a_a$ & 1 &   maximum slope in the astrocytic population\\
$\theta_e$& 4 &  half-maximum firing threshold in the excitatory population\\
$\theta_i$ & 3.7 &  half-maximum firing threshold in the inhibitory population \\
$\theta_a$ & 3 &  half-maximum firing threshold in the astrocytic population \\
$c_{EE}$ & 22 &  self-excitation strength \\
$c_{EI}$ & 14 & excitatory connection strength  \\
$c_{II}$& 0 &self-inhibition \\
$c_{EA}$& 10 & neuronal drive to the astrocytic population\\
$\alpha$& 0.9 & proportion of the external input to the excitatory population \\
$K_p$ & 0.9 & excitatory (external) input  \\
$c_{AE}$& [-5 50] & astrocytic input to the excitatory population\\
$c_{IE}$& [0 40] &  inhibitory connection strength \\
    \end{tabular}
    \caption{Model parameters of the extension of Wilson-Cowan model with astrocytic population.}
    \label{tab:wc_par_astro}
\end{table}

\begin{table}[h!]
\centering 
\begin{tabular}{|c|c|c|} 
\hline
\multicolumn{3}{|c|}{\textbf{Eigenvalues at saddle \#1}} \\
\hline
$0.5152$ & $-0.8747$ & $-1.0131$ \\
\hline 
\multicolumn{3}{|c|}{\textbf{Eigenvalues at saddle \#2}} \\
\hline
$2.4718$ & $-1.2702$ & $-1.8253$  \\
\hline
\end{tabular}
\caption{Eigenvalues associated with saddle nodes in Wilson-Cowan model with astrocytes at the SNIC$^2$ organizing center.}\label{table_eig1star}
\end{table}

\begin{table}[!h]
\centering 
\begin{tabular}{|c|c|c|} 
\hline
\multicolumn{3}{|c|}{\textbf{Eigenvalues at saddle node \#1}} \\
\hline
$7.225 \times10^{-5}$ & $-1.0422$ & $-0.9904$ \\
\hline 
\multicolumn{3}{|c|}{\textbf{Eigenvalues at saddle node \#2}} \\
\hline
$-3.4403 \times 10^{-4}$ & $-1.85$ & $-1.9595$  \\
\hline
\end{tabular}
\caption{Eigenvalues associated with saddle points during bistability in Wilson-Cowan model with astrocytes.}\label{table_eig2star}
\end{table}

\clearpage
\begin{figure*}[!h]  
    \centering
    \includegraphics[width=0.9\textwidth]{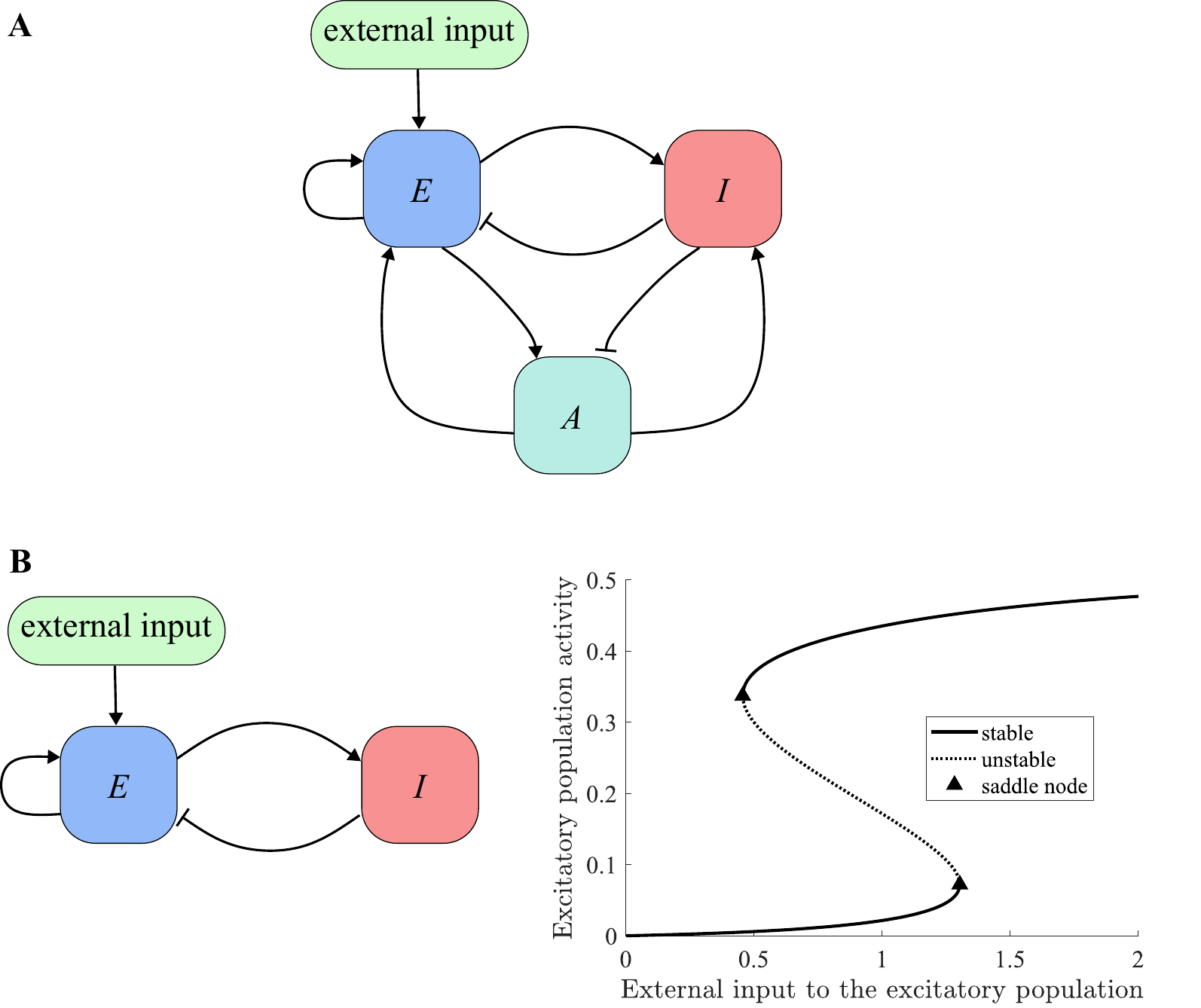}  
    \caption{(\textbf{A})Schematic description of extension of the Wilson-Cowan model with population of astrocytes ($A$), bidirectionally interacting with excitatory ($E$) and inhibitory ($I$) populations. (\textbf{B}) We consider the scenario in which Wilson-Cowan framework without astrocytes does not produce oscillatory dynamics, but transitions between down state, bistability and up state.}
    \label{fig:astro_sup1} 
\end{figure*}
\clearpage
\begin{figure*}[!t]  
    \centering
    \includegraphics[width=0.57\textwidth]{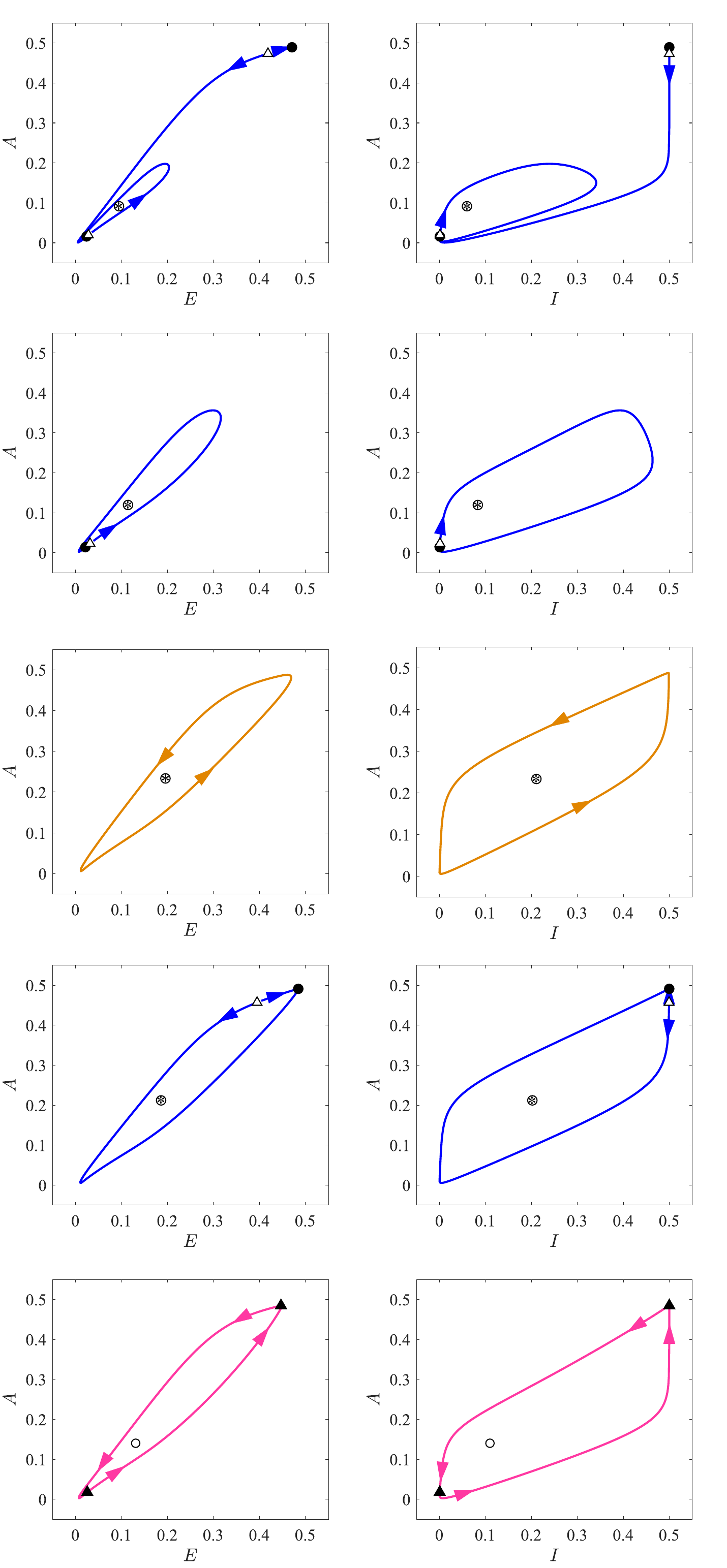}  
    \caption{Phase portraits depicting activity of excitatory population vs astrocytes (left) and  inhibitory population vs astrocytes corresponding to the phase portraits of excitatory vs inhibitory population activity in \autoref{fig:astro} in the main text.}
    \label{fig:astro_sup2} 
\end{figure*}

\clearpage
\subsection*{Jansen-Rit model: higher dimensional model of electrical activity in cortical column}\label{jrmodel}
We demonstrate the consistency of SNIC$^2$ phenomenon in higher-dimensional models of neuronal network dynamics. To this end, we choose the Jansen-Rit model that describes the mean-field dynamics of coupled excitatory and inhibitory neural populations and is often used to study rhythmic and epileptic activity in cortical and thalamic networks \cite{jansen_electroencephalogram_1995}.
\begin{align}
    \frac{dx_1}{dt} &= y_1 \\
    \frac{dx_2}{dt} &= y_2 \\
    \frac{dx_3}{dt} &= y_3 \\
    \frac{dy_1}{dt} &= Aa(I+\sigma(v_{\text{max}_1}, v_{01}, r_1, (c_2x_2 - c_4x_3)))-2ay_1-a^2x_1\\
    \frac{dy_2}{dt} &= Aa\sigma(v_{\text{max}_2}, v_{02}, r_2, (c_1x_1))-2ay_2-a^2x_2\\
    \frac{dy_3}{dt} &= Bb\sigma(v_{\text{max}_3}, v_{03}, r_3, (c_3x_1))-2by_3-b^2x_3, 
\end{align}
The variables $x_1$, $x_2$ and $x_3$ represent the postsynaptic potentials generated by the pyramidal neurons, excitatory interneurons, and inhibitory interneurons, respectively. Their corresponding derivatives are denoted by $y_1$, $y_2$ and $y_3$. The parameters $A$ and $B$ set the maximum size of excitatory and inhibitory postsynaptic potential, respectively. The parameters $a$/$b$ determine how quickly the excitatory/pyramidal and inhibitory responses rise and decay. The $c$ parameters define the strength of the synaptic connections between different neural populations, while $I$ is the external input to the population of the pyramidal neurons. In the canonical version of Jansen-Rit model the external input is provided to the excitatory population as the majority of the input in the system is thought to be integrated via the interneurons. However, in \cite{moran_neural_2007} the case of pyramidal-to-pyramidal neuronal population connections is considered, justifying the external input to the pyramidal population in our model. We also note that we consider the external input to be both positive and negative here as per the Liley model in \cite{bojak_modeling_2005}. The average membrane potential of the neuronal population is then transformed into a firing rate using a sigmoidal function:
\begin{align}
\sigma(v_{\text{max}},v_0,r,x)=\frac{v_{\text{max}}}{1+\exp((v_0-x)r)},
\end{align}
where the parameters $v_{\text{max}}$, $v_0$ and $r$ correspond to maximum threshold, half-maximum activation and reciprocal of the activation slope, respectively.

With the parameter values given in \autoref{tab:jr_par}, a SNIC$^2$ organizing center emerges in the two-parameter bifurcation diagram spanned by the excitatory input $I$ and the inhibitory connection strength ($c_4$) (\autoref{fig:4}, with \autoref{fig:ex4} for the extended version of the two-parameter bifurcation diagram). Strikingly, the bifurcation structures associated with analogous parameters in Wilson-Cowan and Jansen-Rit models closely mirror each other, reflecting how the Jansen–Rit equations act as a higher-dimensional extension of the Wilson–Cowan framework that governs the network’s dynamics. To visualize the model's dynamic behavior we consider 2-dimensional projections in the plane span by the activity of the pyramidal ($y_1$) and inhibitory ($y_3$) neuronal populations. We find bistability characterized by two stable, two saddle and one unstable equilibria (\autoref{fig:4}(\textbf{I})). In our analysis, we found that the saddle points possess one positive eigenvalue, with all other eigenvalues being negative (\autoref{table_eig2}). This means that the unstable manifold is one-dimensional and hence in the phase portraits featuring saddle points we visualize only the one-dimensional unstable manifolds. Staring in the bistable region, decreasing the strength of inhibitory connection leads to the switch to its canonical form of two stable equilibria separated by the saddle point (\autoref{fig:4}(\textbf{Ia})). On the other hand, increasing the strength of the inhibitory connection strength causes the destruction of the upper equilibrium and moves the system to an excitable down state with a unstable, a saddle and a stable point (\autoref{fig:4}(\textbf{II})). By means of increasing the external input (here it changes from being inhibitory to excitatory) the system transitions to oscillatory dynamics via SNIC bifurcation (\autoref{fig:4}(\textbf{III})). In the oscillatory regime, decreasing inhibitory connection strength leads to an oscillatory dynamics loss via another SNIC bifurcation a transition to an excitable up state characterized by an unstable, a saddle, and a stable equilibria (\autoref{fig:4}(\textbf{IV})). As the inhibitory connection strength is decreased even further, the system enters the canonical up state mode (\autoref{fig:4}(\textbf{IVa})). The transitions between these dynamic regimes are governed by a SNIC$^2$ organizng center  (\autoref{fig:4} phase plane at the star point). We note that one eigenvalue of both saddle nodes is zero, while the other five for both equlibria are negative (\autoref{table_eig1} in Appendix). In the Wilson-Cowan and Tsodyks-Markram the non-zero eigenvalues are also negative, giving rise to the stable saddle node manifolds. This further highlights similarity between the two-dimensional models and the higher-dimensional case we consider here.

\begin{figure*}[h]  
    \centering
    \includegraphics[width=0.99\textwidth]{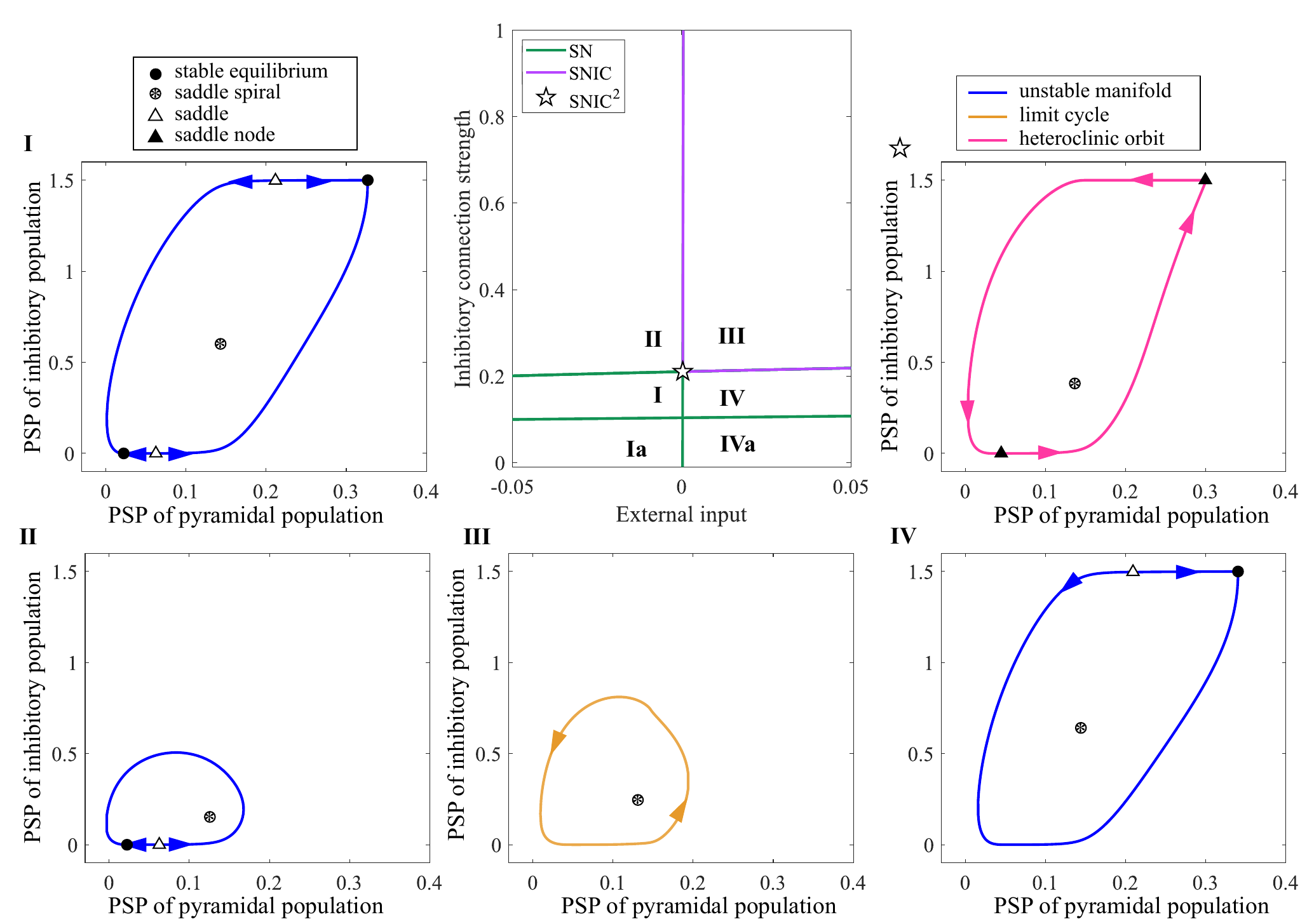}  
    \caption{Visualization of the dynamics in the Jansen-Rit model around SNIC$^2$ organizing center shown in the bifurcation diagram between the parameters responsible for the external input ($I$) and inhibitory connection strength ($c_4$). The bifurcation diagram shows arrangement of the saddle node (SN) and saddle node on invariant circle (SNIC) curves. The phase portraits visualize the regions surrounding SNIC$^2$ organizing center. While regions \textbf{Ia} and \textbf{IVa} are not shown they represent canonical bistability and only up state dynamics.}
    \label{fig:4} 
\end{figure*}

\begin{figure*}[h]  
    \centering
    \includegraphics[width=0.6\textwidth]{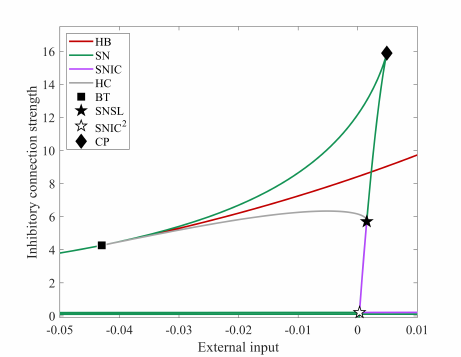}  
    \caption{Extended version of the two-parameter (external input, $I$ and inhibitory connection strength $c_4$) bifurcation diagram in \autoref{fig:4}. The diagram features saddle node (SN), Hopf (HB), saddle node on invariant circle (SNIC), homoclinic (HC) curves. Additionally, there is a Bogdanov-Takens point (BT), a cusp (CP), a saddle node separatrix loop (SNSL), as well as a SNIC$^2$.}
    \label{fig:ex4} 
\end{figure*}

\begin{table} 
    \centering
    \begin{tabular}{c|c|c}
   Parameter & Value & Description\\
    \hline
$A$ &  0.334 & Maximum size of excitatory postsynaptic potential \\
$a$& 1 &  	Time constant of excitatory postsynaptic potential  \\
$B$ & 1.5 &  Maximum size of inhibitory postsynaptic potential  \\
$b$ & 0.5 & 	Time constant of inhibitory postsynaptic potential   \\
$v_{\text{max1}}$& 1 &   Maximum threshold (pyramidal) \\
$v_{\text{max2}}$& 1 &  Maximum threshold (excitatory) \\
$v_{\text{max3}}$& 0.5 &  Maximum threshold (inhibitory) \\
$v_{01}$ & 0.084 & Half-maximum activation (pyramidal)\\
$v_{02}$& 0.084 & Half-maximum activation (excitatory)\\
$v_{03}$& 0.5 & Half-maximum activation (inhibitory)\\
$r_{1}$ & 40 & Reciprocal of the activation slope (pyramidal)\\
$r_{2}$& 40 & Reciprocal of the activation slope (excitatory) \\
$r_{3}$& 30 & Reciprocal of the activation slope (inhibitory) \\
$c_{1}$ & 0.469 & Connection strength parameter from pyramidal to excitatory population\\
$c_{2}$& 1.5 & Connection strength parameter from excitatory to pyramidal population \\
$c_{3}$& 3.4 & Connection strength parameter from pyramidal to inhibitory population \\
$I$& [-0.06, 0.05] & External input \\
$c_{4}$& [0 20] & Connection strength parameter from inhibitory to pyramidal population \\
    \end{tabular}
    \caption{Model parameters of Jansen-Rit model.}
    \label{tab:jr_par}
\end{table}

\begin{table}[h!]
\centering 
\begin{tabular}{|c|c|c|c|c|c|}
\hline 
\multicolumn{6}{|c|}{\textbf{Eigenvalues at saddle \#1}} \\
\hline
$-2.17$ & $-1.00 + 1.17i$ & $-1.00 - 1.17i$ & $0.17$ & $-0.48$ & $-0.52$ \\
\hline
\multicolumn{6}{|c|}{\textbf{Eigenvalues saddle \#2}}  \\
\hline
$-2.60$ & $-1.00 + 1.60i$ & $-1.00 - 1.60i$ & $0.59$ & $-0.47$ & $-0.53$ \\
\hline
\end{tabular}
\caption{Eigenvalues associated with the saddle equilibria of Jansen-Rit model in bistable region.}\label{table_eig2}
\end{table}

\begin{table}[h!]
\centering 
\begin{tabular}{|c|c|c|c|c|c|} 
\hline
\multicolumn{6}{|c|}{\textbf{Eigenvalues at saddle node \#1}} \\
\hline
$1.00 + 1.00i$ & $-1.00 - 1.00i$ & $-2.00$ & $6.89\times10^{-5}$ & $-0.49$ & $-0.51$ \\
\hline 
\multicolumn{6}{|c|}{\textbf{Eigenvalues at saddle node \#2}} \\
\hline
$-1.00 + 1.00i$ & $-1 - 1.00i$ & $1.59\times10^{-5}$ & $-2.00$ & $-0.50$ & $-0.50$ \\
\hline
\end{tabular}
\caption{Eigenvalues associated with saddle nodes in Jansen-Rit model at the SNIC$^2$ organizing center.}\label{table_eig1}
\end{table}

\newpage
\subsection*{Morris-Lecar model: electrical activity of a single neuron} \label{ML}

To explore the presence of SNIC$^2$ beyond the scope of phenomenological mean-field models, we consider Morris-Lecar framework \cite{morris_voltage_1981}, that describes a membrane potential ($V$) of an excitable cell. The membrane potential is tightly related to the activity of the potassium channels. The model captures this dynamics by introducing the gating variable ($N$), which represents the fraction of open potassium channels. The interaction between membrane potential and the gating variable is given by the following system of ODEs, that is taken from \cite{matzakos-karvouniari_biophysical_2019}:
\begin{align}
    \frac{dV}{dt} &= (-g_L(V-V_L)-g_{Ca}M_{ss}(V)(V-V_{Ca})-g_KN(V-V_K)+I)/C \\
    \frac{dN}{dt} &= \tau(V)N_{ss}(V)-N.  
\end{align}
The parameters $g_L$, $g_{Ca}$ and $g_K$ stand for maximum conductance for leak, calcium and potassium conductance, respectively, while equilibrium potential corresponding to leak, calcium and potassium is represented by the parameters $V_L$, $V_{Ca}$ and $V_K$, respectively. The parameter $C$ corresponds to membrane capacity, while the parameter $I$ corresponds to the external input to the cell. The functions $M_{ss}$, $N_{ss}$ and $\tau$ are the auxiliary functions used to describe voltage-dependent behavior of ion channels. The functions $M_{ss}$ and $N_{ss}$ are representative of the steady-state activation of calcium and potassium channels, respectively, and both have sigmoidal shape:
\begin{align}
    M_{ss}(V)&=\frac{1}{2}(1+\text{tanh}(\frac{V-V_1}{V_2})) \\
    N_{ss}(V)&=\frac{1}{2}(1+\text{tanh}(\frac{V-V_3}{V_4})), 
\end{align}
where $V_1$ and $V_3$ represent the half-maxima of the respective functions, while $V_2$ and $V_4$ stand for their slopes. Since potassium channels take time to open or close, $\tau$ represents the voltage-dependent time constant that governs this process and is given as: 
\begin{align}
    \tau(V)=\text{cosh}(\frac{V-V_3}{V_4}). 
\end{align}
While the Morris-Lecar model captures the biophysics of a single neuron, it exhibits similar transitions membrane potential dynamics as mean-field population models like Wilson-Cowan, Tsodyks-Markram, and Jansen-Rit. Using the parameter values in \autoref{tab:ml_par}, we identify the presence of SNIC$^2$ under the change of external input and maximum potassium conductance  (\autoref{fig:5}). We identify a bistable state with two stable, two saddle and one unstable equilibrium (\autoref{fig:5}(\textbf{I})). Here the bistability allows the system to switch between hyperpolarized and depolarized states. Decreasing maximum conductance or external input drives the system toward a canonical bistable state, characterized by two stable fixed points separated by a saddle (\autoref{fig:5}(\textbf{Ia})). Whereas increase in maximum potassium conductance leads to a saddle node bifurcation that eliminates the depolarized state, leaving only hyperpolarized state present with a stable, a saddle and an unstable equilibria, allowing for a long transient excursion subject to a suprathreshold perturbation (\autoref{fig:5}(\textbf{II})). From this state we can transition to the canonical hyperpolarized state with only one stable equilibrium via decreasing the external input (\autoref{fig:5}(\textbf{IIa})). On the other hand, raising external input puts the system through the SNIC bifurcation, giving rise to the limit cycle (\autoref{fig:5}(\textbf{III})). From there, decrease of the maximum conductance drives the system to the loss of oscillatory dynamics and the birth of the depolarized state with a saddle, an unstable and a stable equilibrium (\autoref{fig:5}(\textbf{IV})). Excessive decrease of the potassium conductance results in a canonical depolarized state with one stable equilibrium (\autoref{fig:5}(\textbf{IVa})). 

\begin{table}[h]
    \centering
    \begin{tabular}{c|c|c}
   Parameter & Value & Description\\
    \hline
$C$ &  1 & Membrane capacity\\
$g_L$ & 2 &   Maximum conductance for leak\\
$g_{Ca}$& 15 &  Maximum conductance for calcium \\
$V_L$ & -60 &  Leak equilibrium potential \\
$V_{Ca}$ & 127 &  Calcium equilibrium potential \\
$V_K$ & -90 & Potassium equilibrium potential  \\
$V_1$& -1.23 &  Half-maximum for steady-state activation of calcium \\
$V_2$& 20 & Slope for steady-state activation of calcium \\
$V_3$& -20 & Half-maximum for steady-state activation of potassium \\
$V_4$ & 7 & Slope for steady-state activation of potassium  \\
$I$& [-100, 50] & External input \\
$g_K$& [0,40] & Maximum conductance for potassium \\
    \end{tabular}
    \caption{Model parameters of Morris-Lecar model.}
    \label{tab:ml_par}
\end{table}

\begin{figure*}   
    \centering
    \includegraphics[width=0.99\textwidth]{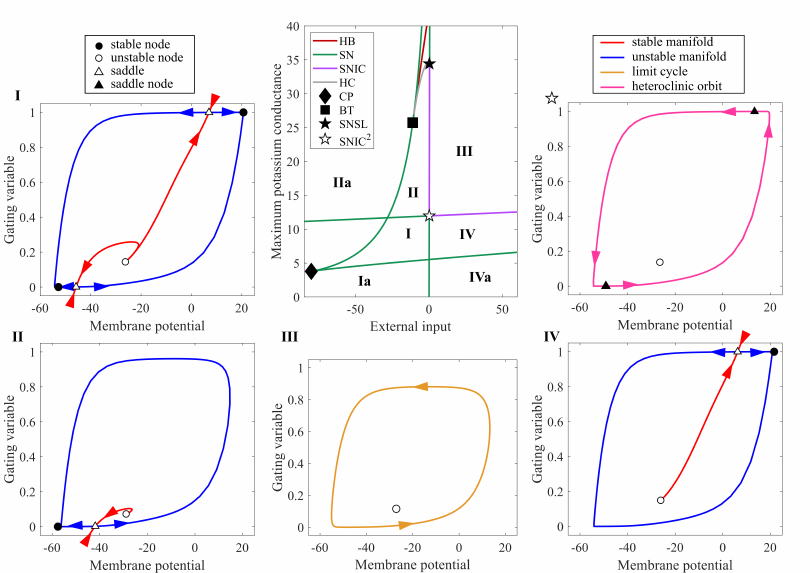}  
    \caption{Visualization of the dynamics in the Morris-Lecar model around SNIC$^2$ organizing center shown in the bifurcation diagram between the parameters responsible for the external input ($I$) and maximum potassium conductance ($g_K$). The bifurcation diagram shows arrangement of the saddle node (SN), Hopf (HB), saddle node on invariant circle (SNIC), and homoclinic (HC) curves. The points BT, CP and SNSL correspond to the Bogdanov-Takens point, cusp, and saddle node separatrix loop, respectively. The phase portraits demonstrate the dynamics in the regions surrounding SNIC$^2$ organizing center. The dynamics of \textbf{Ia}, \textbf{IIa} and \textbf{IVa} is not shown but corresponds to the canonical bistability, only down and only up states, respectively.}
    \label{fig:5} 
\end{figure*}

\fi

\end{document}